\documentclass[11pt,a4paper]{article}
\pdfoutput=1
\usepackage{jheppub}
\usepackage{amsmath}%
\usepackage{amsfonts}%
\usepackage{amssymb}%
\usepackage{graphicx}
\usepackage{slashed}
\usepackage{comment}

\newcommand{\be}{\begin{equation}}
\newcommand{\ee}{\end{equation}}

\title{Gravitational Effects of Disformal Couplings}

\author[a]{Philippe Brax}
 \author[b]{ and Anne-Christine Davis}

\affiliation[a]{Institut de Physique Th\'eorique, Universit\'e Paris-Saclay, CEA,CNRS,\\
F-91191Gif sur Yvette, France }
\affiliation[b]{DAMTP, Centre for Mathematical Sciences, University of Cambridge,
  CB3 0WA, UK}
	
	\emailAdd{philippe.brax@cea.fr}
\emailAdd{a.c.davis@damtp.cam.ac.uk}

\abstract{We consider how a nearly massless scalar field conformally and disformally coupled to matter can affect the dynamics of gravitationally interacting bodies.   We focus on  the case of two interacting objects and  we obtain the effective metric driving the dynamics of the two body system when reduced to one body in the centre of mass frame. We then concentrate  on the case of a light particle
in the scalar and gravitational fields generated by a heavy object and find the effects of the conformal and disformal couplings on the body's trajectory such as
the advance of perihelion and the Shapiro time delay. The disformal coupling leads to a negligible  contribution to the Shapiro effect and therefore no constraint from the Cassini experiment. On the other hand, it contributes
to the perihelion advance leading to a weak bound on the strength of the disformal coupling itself. Finally, we remark that the disformal coupling gives rise to a contribution to the perihelion advance which varies quadratically with the mass of the heavy body, leading to  possible strong effects for stars in the vicinity of astrophysical black holes. For neutron stars in a binary system, the disformal effects  vary as the quartic power of the size of the orbit which might lead to interesting consequences in the inspiralling phase prior to a merger.  }

\begin{document}

\maketitle
\flushbottom

\section{Introduction}
Nearly massless scalar fields are ubiquitous in cosmology \cite{Ratra:1987rm,Wetterich:1987fm,Copeland:2006wr}. They could play a role in generating the late time acceleration of the expansion of the Universe. They could also belong to an
extended gravitational sector of the theory describing the Universe \cite{Sotiriou:2008rp,Khoury:2013tda,deRham:2012az,deRham:2014zqa}. In this work we shall consider that such a scalar could be both conformally and disformally coupled to matter \cite{Bekenstein:1992pj}. The effects of a conformal
coupling are well known \cite{Damour:1992we,Julie:2017pkb} and must be suppressed in the solar system in order to comply with gravitational tests such as the ones performed by the Cassini probe \cite{Bertotti:2003rm} (existence of a fifth force) or the Lunar Laser Ranging experiment (test of the strong equivalence principle in the earth-moon-sun system) \cite{Williams:2004qba}. The resulting bounds on the coupling $\beta$ are severe and screening mechanisms have been invoked in order to comply with rather unnaturally small values of $\beta$ \cite{Khoury:2003aq,Damour:1994zq,Vainshtein:1972sx,Babichev:2009ee}. Another type of interaction, the disformal coupling, could also play a role in the interactions between matter and the scalar field. This coupling has been constrained using numerous probes \cite{Koivisto:2008ak,Zumalacarregui:2010wj,Koivisto:2012za,vandeBruck:2013yxa,Brax:2013nsa,Neveu:2014vua,Sakstein:2014isa,Sakstein:2014aca,Ip:2015qsa,Sakstein:2015jca,vandeBruck:2015ida,vandeBruck:2016cnh}. It can influence the dynamics of compact bodies as a one loop effective interaction similar to the Casimir-Polder force can be generated between such objects \cite{Kaloper:2003yf}. It can also have effects on the atomic energy levels or even the burning rate of stars in astrophysics \cite{Brax:2014vva}. Finally, as a four-body interaction, it can be tested at accelerators such as the LHC \cite{Brax:2015hma}. In this paper, we will focus on the gravitational physics of such a disformal coupling \cite{Ip:2015qsa,Sakstein:2015jca}, in conjunction with a conformal coupling, in the presence of celestial bodies. We will derive an effective one body metric which describes the dynamics of two such interacting bodies at leading order in $G_N$. This will allow us to consider the disformal effects on the classical tests such as the advance of perihelion or the time delay of radio-wave signals. The effective one body metric may also eventually allow us to consider the inspiralling emission of gravitational wave by two rotating bodies, although we leave it for further work.

We find that the Shapiro time delay as probed by the Cassini experiment does not depend (or extremely weakly) on the disformal coupling. On the other hand, the perihelion advance of a light body in the presence of a heavy object is non vanishing. We find that it varies quadratically with the mass of the heavy object and quartically with the size of the orbit. This may have consequences for the dynamics of stars in the vicinity of astrophysical black holes \cite{2017ApJparsa} or during the inspiralling phase of neutron star mergers.

In section \ref{sec:disf} we study the solutions to the Klein-Gordon equation involving a conformal and disformal coupling to matter for point sources. In section \ref{sec:dyn2} we consider the case of two interacting bodies whilst  in section \ref{sec:dyn1} we consider the dynamics of a light particle close to a heavy body. We discuss possible consequences for the dynamics of stars  close to astrophysical black holes and binary systems of neutron stars  in section \ref{sec:con}.

\section{Disformal radiation}
\label{sec:disf}
\subsection{Ladder expansion}

In this section we consider the scalar emission from a moving body when the coupling between matter and the scalar field is mediated by the metric
\be
g_{\mu\nu}= A^2(\phi) g_{\mu\nu}^E+ \frac{2}{M^4} \partial_\mu\phi \partial_\nu \phi
\ee
where the conformal factor
\be
A(\phi)= e^{\beta \phi/m_{\rm Pl}}
\ee
is characterised by the constant coupling $\beta$ and the disformal interaction is specified by the suppression scale $M$.
We could have chosen more complex function \cite{Brax:2010kv} such as a quadratic function $A(\phi)$, e.g. as for the environmentally dependent dilaton \cite{Brax:2010gi} and symmetron \cite{Hinterbichler:2010es}. Here we consider the simplest case of a field independent coupling $\beta$. Similarly the disformal part could be more complex with $1/M^4 \to B(\phi, (\partial \phi)^2)$. In the following we focus on the simplest case where the disformal coupling depends only on the constant coupling scale $M$.
Matter couples minimally to $g_{\mu\nu}$ such that the total action reads
\be
S=\int d^4x \sqrt{-g_E}\left ( \frac{R_E}{16\pi G_N}-\frac{1}{2} (\partial\phi)^2 -V(\phi)\right ) +S_m(\psi_i, g_{\mu\nu})
\ee
in the Einstein frame for the Einstein-Hilbert action and we have introduced a potential $V(\phi)$ for the scalar field. The matter fields are denoted by $\psi_i$ and their action is $S_m$.
In the following, we will focus of nearly massless scalar and take $V(\phi)=0$. We could have considered the case of screened models with a non-trivial $V(\phi)$ \cite{Joyce:2014kja}. Effectively in this case and in a given environment such as the solar system, the mass of the scalar field between the sun and the planets is small, i.e. the scalar field is not Yukawa-screened, and  in the screened models with either the chameleon or the Damour-Polyakov screenings the scalar charge of these objects $\beta_{\rm eff}$ is reduced to pass the solar system tests such as the Cassini bound \cite{Bertotti:2003rm}. We leave a detailled analysis of screened models for the future and concentrate on the
case of a massless field with a small coupling $\beta$.

The gravitational dynamics are dictated by the Einstein equation
\be
R_{\mu\nu}-\frac{1}{2} R g_{\mu\nu}= 8\pi G_N (T_{\mu\nu}+ T^\phi_{\mu\nu})
\ee
where the matter energy-momentum tensor is
\be
T_{\mu\nu}=-\frac{2}{\sqrt{-g^E}} \frac{\delta S_m}{\delta g^{\mu\nu}_E}
\ee
and the corresponding one for the scalar field is
\be
T^\phi_{\mu\nu}= \partial_\mu\phi \partial_\nu \phi-\frac{(\partial\phi)^2}{2} g_{\mu\nu}^E.
\ee
The dynamics of the scalar field are given by the Klein-Gordon equation
\be
\Box \phi= -\beta\frac{T}{m_{\rm Pl}} +\frac{1}{M^4} D_\mu( A^{-2}(\phi)  \partial_\nu\phi T^{\mu\nu})
\ee
where $D_\mu$ is the covariant derivative for the Einstein metric.
The Bianchi identity implies the non-conservation equation
\be
D^\mu T_{\mu\nu}= \frac{\beta T}{m_{\rm Pl}} \partial_\nu \phi - \frac{1}{M^4} D_\mu( A^{-2}(\phi)\partial_\lambda\phi  T^{\mu\lambda}) \partial_\nu \phi.
\ee
In the following we will be interested in the leading terms at the $1/m_{\rm Pl}$ order in $\phi$. Indeed this leads to contributions to the
interaction potential between two objects in $\beta \phi /m_{\rm Pl}$ proportional to $G_N$. As we are only focussing on the leading $G_N$ contributions to the
dynamics of interacting bodies and to leading order in $1/M^4$, we can safely consider that the matter energy momentum is conserved
\be
D^\mu T_{\mu\nu}=0
\ee
at this order leading to the Klein-Gordon equation
\be
\Box \phi= -\beta\frac{T}{m_{\rm Pl}} +\frac{1}{M^4}   D_\mu \partial_\nu\phi T^{\mu\nu}.
\ee
The Klein-Gordon equation can be solved iteratively as
\be
\phi= \phi^{(0)}+ \delta\phi
\ee
where
\be
\Box \phi^{(0)}= -\beta\frac{T}{m_{\rm Pl}}
\ee
is non-trivial when $\beta\ne 0$
and
\be
\Box \delta \phi-\frac{1}{M^4} D_\mu \partial_\nu\delta \phi T^{\mu\nu}=\frac{1}{M^4} D_\mu \partial_\nu\phi^{(0)} T^{\mu\nu}.
\ee
Defining the retarded propagator $G$ as
\be
\Box G (x,x')= \delta^{(4)} (x-x')
\ee
we have
\be
\phi^{(0)}(x)= -\frac{\beta}{m_{\rm Pl}}\int d^4 x' G(x-x') T(x')
\ee
and we can find a series representation of the solution  corresponding to an expansion in  ladder diagrams
\be
\delta \phi= \sum_{n\ge 0} \delta \phi^{(n)}
\ee
where
\be
\Box \delta \phi^{(0)}=\frac{1}{M^4} D_\mu \partial_\nu\phi^{(0)} T^{\mu\nu}.
\ee
and
\be
\Box \delta \phi^{(n+1)}=\frac{1}{M^4} D_\mu \partial_\nu\delta\phi^{(n)} T^{\mu\nu}.
\ee
This implies  that
\be
\delta\phi^{(0)}= \frac{1}{M^4} \int d^4 x' G(x-x') D_\mu\partial_\nu \phi^{(0)}(x') T^{\mu\nu}(x')
\label{first}
\ee
and
\be
\delta \phi^{(n+1)}(x)= \frac{1}{M^4} \int d^4x' G(x-x') D_\mu\partial_\nu \delta\phi^{(n)}(x') T^{\mu\nu}(x').
\ee
Each iteration brings in another insertion of the energy-momentum tensor and is suppressed by a higher power of $M^4$. Hence to be consistent with our approximation we only consider the first two steps.
Notice that the solution vanishes in the absence of a conformal coupling $\beta$.

{ We have neglected the possible effects coming from the cosmological background density. When the matter system is embedded in the cosmological background
with an energy-momentum $T^{\mu\nu}_{\rm cosmo}$, one can separate the scalar field as $\phi= \phi_{\rm cosmo}+ \bar \phi$ where
\be
\Box \phi_{\rm cosmo}= -\beta\frac{T_{\rm cosmo}}{m_{\rm Pl}} +\frac{1}{M^4}   D_\mu \partial_\nu\phi_{\rm cosmo} T^{\mu\nu}_{\rm cosmo}
\ee
and the background metric is now of the Friedmann-Lemaitre-Robertson-Walker type. The local matter density generates a scalar field such that
\be
\Box \bar \phi= -\beta\frac{T}{m_{\rm Pl}} +\frac{1}{M^4}   D_\mu \partial_\nu\bar \phi T^{\mu\nu} + \frac{1}{M^4}   D_\mu \partial_\nu\bar \phi T^{\mu\nu}_{\rm cosmo}+\frac{1}{M^4}   D_\mu \partial_\nu \phi_{\rm cosmo} T^{\mu\nu}.
\ee
There are two new source terms which involve the cosmological energy-momentum tensor and the derivatives of the cosmological solution. As the cosmological matter density is negligible compared
to the matter density in the moving objects we are considering and the variation of the cosmological solution is on time scales much larger than the rapid motion of the moving bodies, we can safely neglect the new source terms. Within this quasi-static approximation, the only effect of the cosmological background is to add to the solution generated by the local matter density  a slowly varying background scalar field $\phi_{\rm cosmo}$.
}

\subsection{Point source}

We now focus on a point source of mass $m$ whose energy momentum tensor reads
\be
T^{\mu\nu} = m  \int d\tau A(\phi) u^\mu u^\nu \delta^{(4)} (x^\mu -x^\mu(\tau))
\ee
where $\tau$ is the proper time of the particle in the Einstein frame such that
\be
u^\mu=\frac{dx^\mu}{d\tau}
\ee
and $u^\mu u_\mu=-1$. Notice that, as we work in the Einstein frame, the mass of the particle becomes $m A(\phi)$ which is field dependent. We will work in the case where
$\beta \phi/m_{\rm Pl} \ll 1$ which will be valid as long as $\beta={\cal O}(1)$ as,   to leading order,  $\beta \phi/m_{\rm Pl} \sim 2\beta^2 \Phi_N$ for an object with Newtonian potential $\Phi_N$. For the objects that
we consider such as the sun $\Phi_N\ll 1$ and we can therefore omit the $A(\phi)\sim 1$ in the mass. { When the cosmological background is taken into account, at leading order, one can keep
track of the effects of the slow variation of the background scalar field by retaining the slow time variation of the mass of the particles using $m\to A(\phi_{\rm cosmo}) m$. }

We will be interested on the effects of a point source on the geometry of space at the leading $G_N$ order and its consequences on the
effective geometry driving the motion of interacting particles. Hence it is sufficient to consider the evolution of the
point particle in Minkowski space and contract the tensors with $\eta_{\mu\nu}$.
Defining the velocity
\be
v^i= \frac{dx^i}{dx^0}
\ee
for the particle, we find that
\be
T=- m \sqrt{1-\vec v^2} \delta^{(3)}( x^i- x^i(x^0))
\ee
where we have $x^0=x^0(\tau)$. Hence an ultra-relativistic particle has a traceless energy momentum, in agreement with the traceless of $T^{\mu\nu}$ for a relativistic fluid.
Using the Green's function in Minkowski space
\be
G(x,x')\equiv G(x-x')=-\frac{1}{2\pi} \theta( x_0-x_0') \delta (( x-x')^2)
\ee
we find that
\be
\phi^{(0)}(x)= -\frac{\beta m}{4\pi m_{\rm Pl}} \frac{1}{1- \vec v. \vec n'}\frac{\sqrt{1-\vec v^2}}{\vert \vec x -\vec x (x'^0)\vert}
\ee
which matches the usual solution of the Poisson equation for a static scalar field sourced by a static point particle.
We have defined the unit vector
\be
\vec n'= \frac{\vec x -\vec x(x'^0)}{\vert \vec x- \vec x(x'^0)\vert}
\ee
and similarly
\be
\vec n= \frac{\vec x -\vec x(x^0)}{\vert \vec x- \vec x(x^0)\vert}.
\ee
Here we have introduced the retarded time
\be
x'^0= x^0-\vert \vec x -\vec x (x'^0)\vert.
\ee
In the same vein, the first iteration of the ladder expansion reads
\be
\delta \phi^{(0)}(x)= -\frac{m}{4\pi M^4\gamma }\frac{\partial_\mu \partial_\nu \phi^{(0)} u^\mu u^\nu}{\vert \vert \vec x-\vec x (x'^0)\vert -\vec v. (\vec x -\vec x (x'^0))\vert}
\ee
where $u^\mu=\gamma(1, v^i)$ and  $\gamma=1/\sqrt{1-\vec v^2}$.
The higher order terms can be deduced by iteration.

For small velocities, we can expand
\be
x^0-x'^0= (1+\delta) \vert \vec x- \vec x(x^0)\vert
\ee
where
\be
\delta \sim  \vec n.\vec v +\frac{\vec v^2}{2} +\frac{(\vec n.\vec v)^2}{2}
\ee
at second order in the velocity and the scalar field becomes
\be
\phi^{(0)}(x)= -\frac{\beta m}{4\pi m_{\rm Pl}}\frac{1-\frac{\vec v^2}{2} +\frac{\vec v_\perp^2}{2}}{\vert \vec x- \vec x(x^0)\vert},
\ee
where we have defined the projection of the velocity in the direction perpendicular to $\vec n$ as
\be
\vec v_\perp= \vec v - (\vec v.\vec n) \vec n.
\ee
This result can also be deduced using Lorentz invariance. In the frame where the particle is static, the solution is
$-\frac{\beta m}{4\pi m_{\rm Pl}}\frac{1}{\vert \vec x'- \vec x'(x^0)\vert}$
where the distance in the static frame $\vert \vec x'- \vec x'(x^0)\vert$  is longer by a factor $(1+ \frac{(\vec v.\vec n)^2}{2})$. Notice that we always
work at the $\vec v^2 $ order as this is enough to deduce the form of the effective metric between two bodies in the leading $G_N$ approximation.

As a side result, we obtain the Green's function for the spatial Klein-Gordon equation in the presence of a slowly moving particle
\be
\Box G_0 (x)= \delta^{(3)} (x^i- x^i(x^0))
\ee
which is given by
\be
G_0(x)= -\frac{1}{4\pi }\frac{1+\frac{\vec v_\perp^2}{2}}{\vert \vec x- \vec x(x^0)\vert}.
\label{G0}
\ee
This will be useful when solving for the Newtonian potential.

At leading order we have for the derivatives of the scalar field
\be
\partial_0 \phi^{(0)}= -\frac{\beta \gamma m}{4\pi m_{\rm Pl}}\frac{\vec v.(\vec x-\vec x(x^0))}{\vert \vec x-\vec x(x^0)\vert^3}
\ee
and
\be
\partial_i \phi^{(0)}= \frac{\beta m}{4\pi \gamma m_{\rm Pl}} \frac{(x^i-x^i(x^0))}{\vert \vec x-\vec x (x^0)\vert^3}
\ee
which should also depend on the acceleration $a^i=\frac{dv^i}{dx^0}$. In the following we will use the fact that the acceleration involves one power of $G_N$ and therefore these terms appear at higher order
in the $G_N$ expansion, i.e. with the approximation that the acceleration
\be
\vec a_A=-\frac{G_N m_A(\vec x -\vec x_A)}{\vert \vec x- \vec x_A\vert^3}
\ee
is of order $G_N$ and induces corrections in $G_N^2$ that we have neglected.
For the velocity dependent part we find that
\be
\delta \phi^{(0)}= 0
\ee
explicitly when only one particle is involved. This is also a result which follows from Lorentz invariance as in the frame where the particle is static, the solution $\phi^{(0)}$ is time independent.
For two bodies the solution does not vanish and will be given below.

\section{Two Body system}
\label{sec:dyn2}

\subsection{The scalar field of moving particles}

When two moving bodies are present, the solution to the Klein-Gordon equation at leading order in $1/M^4$ can be obtained in two steps.
The first steps consists in solving
\be
\Box \phi^{(0)}= -\beta\frac{T^A +T^B}{m_{\rm Pl}}
\ee
where the energy momentum tensor contains both the parts from particles $A$ and $B$.
The solution is simply given by the linear combination
\be
\phi^{(0)}= \phi^{(0)}_A+ \phi^{(0)}_B
\ee
where we have
\be
\phi^{(0)}_{A,B}( x)= -\frac{\beta m_{A,B}(1-\frac{\vec v_{A,B}^2}{2}+\frac{\vec v_{A,B\perp}^2}{2})}{4\pi m_{\rm Pl}\vert \vec x-\vec x_{A,B}\vert}.
\label{corB}
\ee
This solution sources the next step in the iteration process where
\be
\phi(x)= \phi^{(0)}(x) +\delta \phi^{(0)}(x)
\ee
is given by
\be
\delta\phi^{(0)}(x)= \frac{1}{M^4} \int d^4 x' G(x-x') \partial_\mu\partial_\nu (\phi_A^{(0)}(x')+\phi_B^{(0)}(x'))( T_A^{\mu\nu}(x')+T_B^{\mu\nu}(x')).
\ee
This leads to four contributions
\be
\delta \phi^{(0)}_{\alpha\beta}(x)= -\frac{m_\alpha}{4\pi M^4\gamma }\frac{\partial_\mu \partial_\nu \phi_\beta^{(0)}(x_\alpha) u_\alpha^\mu u_\alpha^\nu}{ \vert \vec x-\vec x_\alpha\vert}
\ee
where $\alpha,\beta=A,B$.
Notice that here the $\vec v_{A,B}^2$ and $\vec v^2_{A,B\perp}$  corrections in (\ref{corB}) are negligible as we neglect the quartic terms in the velocities.
It turns out then that $\delta \phi^{(0)}_{AA}$ and $\delta \phi^{(0)}_{BB}$ both vanish whilst
\begin{eqnarray}
&&\delta \phi^{(0)}_{AB}=-\frac{\beta m_A m_B}{16\pi^2 m_{\rm Pl}} \frac{(\vec v_A -\vec v_B)^2 -3 (\vec n_{AB}.(\vec v_A-\vec v_B))^2}{M^4\vert \vec x-\vec x_A\vert \vert \vec x_B-\vec x_A\vert^3}\nonumber \\
&&\delta \phi^{(0)}_{BA}=-\frac{\beta m_A m_B}{16\pi^2 m_{\rm Pl}} \frac{(\vec v_A -\vec v_B)^2 -3 (\vec n_{AB}.(\vec v_A-\vec v_B))^2}{M^4\vert \vec x-\vec x_B\vert \vert \vec x_B-\vec x_A\vert^3}\nonumber \\
\label{corA}
\end{eqnarray}
where $\vec n_{AB}$ is the unit vector between $A$ and $B$ .
Notice that this involves only the Galilean invariant combination $(\vec v_A -\vec v_B)$.

\subsection{The  gravitational fields of moving particles}
We now consider the interaction between two particles A and B which are conformally and disformally coupled to the scalar field. We shall work to leading order in $G_N$ and $1/M^4$ and in the non-relativistic
limit where $\vec v_{A,B}^2 \ll 1$.
The action for the two bodies can be obtained using
\be
S= -m_A\int d\tau_A \sqrt{-g_{\mu\nu}^B u^\mu_A u^\nu_A}-m_B\int d\tau_B \sqrt{-g_{\mu\nu}^A u^\mu_B u^\nu_B}+ \delta S_{AB}
\ee
where the correction term $\delta S_{AB}$ comes from the evaluation of the field action, both from General Relativity and its  scalar counterpart.

This calculation can be performed in a number of ways but we shall find convenient to work in the non-relativistic limit of GR \cite{Kol:2007bc,Kol:2010si}. To do so, let us decompose the Einstein metric according to
\be
ds^2= -e^{2\Phi_N} (dt-A_idx^i)^2 + e^{-2\Phi_N} \gamma_{ij} dx^i dx^j
\ee
where $\Phi_N$ is the Newtonian potential and $A_i$ is responsible for gravi-magnetism. We have chosen to treat the spatial metric $\gamma_{ij}=\delta_{ij}$ as flat.
In this gauge, the Einstein-Hilbert action can be written as
\be
S_{EH}=-\frac{1}{16\pi G_N} \int d^4x \left ( 2 (D_i\Phi_N)^2 -\frac{e^{2\Phi_N}}{4} F^2 + 4 \dot A^i D_j \Phi_N +6 \dot \Phi_N e^{-4\Phi_N}\right)
\ee
in a $(3+1)$ decomposition of the Kaluza-Klein type. The covariant derivative is $D_i=\partial_i + A_i \partial_t$ and the field strength $F_{ij}= \partial_i A_j -\partial_j A_i$ where
indices are raised and lowered with the flat $\delta_{ij}$.
The Einstein-Hilbert action must be complemented with a gauge fixing action which imposes the harmonic gauge $\partial_\nu(\sqrt{-g} g^{\mu\nu})=0$ and reads now
\be
S_{GF}= \frac{1}{32\pi G_N} \int d^4x \left ( (e^{2\Phi_N} D^i A_i +4 e^{-2\Phi_N} \dot \Phi_N)^2 - \dot A_i^2\right).
\ee
The equations of motion are then
\be
\Box \Phi_N=   4\pi G_N (T^{00}+ T^i_i)
\ee
for the Poisson equation and
\be
\Box  A_i= 16\pi G_N T^{0i}
\ee
for the Maxwell equation of gravi-magnetism.
The Newtonian potential is modified compared to the static case  by the fact that the sources are  moving compared to a nearly-Minkowski background. As a result, distances are effectively contracted by special
relativistic effects.
The solution to the Poisson equation for a single moving source of velocity $\vec v_A$ reads
\be
\Phi_N ( x)= -\frac{G_N m_A}{\vert \vec x-\vec x_A\vert} (1+\frac{3}{2} \vec v_A^2 +\frac{\vec v^2_{A,\perp}}{2})
\ee
where the correction factor comes from the fact that $T^i_i$ brings one factor of $\vec v_A^2$. Another factor of $\vec v_A^2/2$ comes from the time dilation factor $d\tau_A= \gamma^{-1}dx^0 =(\sqrt{1-\vec v_A^2})dx^0$ between proper time
and background time. Finally the Klein-Gordon equation must be solved with a spatial Dirac function as a source. We have already obtained this solution in the form of the Green's function $G_0$, i.e.  \ref{G0}.
Here we have introduced
\be
\vec n= \frac{\vec x-\vec x_A}{\vert \vec x-\vec x_A\vert}
\ee
as the unit vector pointing towards $\vec x$ from $\vec x_A$ and defined the perpendicular velocity
\be
\vec v_\perp= \vec v- (\vec n.v) \vec n
\ee
such that $\vec v^2_\perp= \vec v^2 - (\vec v.\vec n)^2$.
This result is nothing but the Newtonian potential after a boost as Lorentz invariance is preserved by the harmonic gauge \cite{Kopeikin:2005vf}.
Similarly the vector field  is given by
\be
A^i= -4\frac{G_N m_A v_A^i}{\vert \vec x-\vec x_A\vert}
\ee
which is again the result of boosting the static Newtonian metric \cite{Kopeikin:2005vf}.

\subsection{The two body action}

\subsubsection{The gravitational action}

The previous solutions to the field equations contribute to the gravitational action and can be expressed as a function of the velocities of the two moving bodies and their positions.
Denoting by
\be
g_{\mu\nu}^E= \eta_{\mu\nu} + h_{\mu\nu}
\ee
the Einstein metric, the gravitational action comprising both the Einstein-Hilbert term and the gauge fixing leads to
\be
S_{EH}+ S_{GF}= -\frac{1}{4} \int d^4x h_{\mu\nu} T^{\mu\nu}.
\ee
Removing the infinite self-energies, the action for interacting particles is obtained as
\be
S_{EH}+ S_{GF}=-\frac{1}{4} \int d^4x \left ( h^A_{\mu\nu} T^{\mu\nu}_B+h^B_{\mu\nu} T^{\mu\nu}_A\right )
\ee
where $h^{A,B}_{\mu\nu}$ is the field generated by $A$ (respectively $B$).
It is useful to notice the identity (up to acceleration terms which are of higher order in $G_N$)
\be
\frac{d}{dt}(\vec n_{AB}. \vec v_B)=\frac{\vec v_{B\perp}^2 -\vec v_{A\perp}.\vec v_{B\perp}}{\vert \vec x_B-\vec x_A\vert}\equiv 0
\label{trick}
\ee
where $\vec n_{AB}$ is the unit vector between $A$ and $B$ and we have $\vec v_A^\perp= \vec v_A - (\vec v_A.\vec n_{AB}) \vec n_{AB}$ (similarly for $\vec v_B^\perp$). The last equality is to be taken as integrated in an action where total derivatives are irrelevant.
The corresponding gravitational Lagrangian becomes
\be
{\cal L}_{EH}+ {\cal L}_{GF}= -\frac{G_N m_A m_B}{\vert x_B-x_A\vert}(1+ \frac{3}{2}(\vec v_A^2+\vec v_B^2) - 4\vec v_A.\vec v_B +\frac{1}{2} \vec v_{A\perp}.\vec v_{B\perp})
\ee
These terms appear as counter-terms to avoid any double-counting in the overall action including matter.

\subsubsection{The matter action}

We can now expand the matter action to second order in the velocity field and get
the Lagrangian for particle $A$ in the fields generated by particle $B$
\be
{\cal L}_A= -m_A
 e^{\Phi_N^B( x_A) } A( \bar \phi^{}_B ( x_A)) \sqrt{1 -2 A_i^B v^i_A - e^{-4\Phi_N^B(x_A) } \vec v_A^2 -\frac{2}{M^4}D_A}
\ee
and symmetrically for particle B. We  will use explicitly  the fact that
\be
\partial_0  \phi^{(0)}_B( x)=-\vec \partial \phi^{(0)}_B( x).\vec v_B.
\ee
and we will denote by
\be
D_A=(\vec\partial \phi^{(0)}_B( x_A) .\vec v_B)^2+ (\vec \partial \phi^{(0)}_B( x_A) . \vec v_A)^2- 2(\vec \partial \phi^{(0)}_B(x_A) . \vec v_A)(\vec \partial \phi^{(0)}_B( x_A) . \vec v_B).
\ee
the part of the action which comes from the disformal term of the metric.
 The scalar field in this action is $\bar \phi^{}_B ( x)$ where the divergent self-energy contributions have been removed at $\vec x=\vec x_A$, i.e.
\be
\phi(x)= \bar \phi_B(x) + {\cal}{O}(\frac{1}{\vert \vec x-\vec x_A\vert}).
\ee
where explicitly
\be
\bar\phi_B (x)= \phi^{(0)}_B(x) +\delta \phi^{(0)}_{BA}(x)
\ee
is the field generated by the particle $B$ evaluated at particle $A$. Notice that there is a component $\delta \phi^{(0)}_{BA}(x)$ which comes from the back-reaction on the scalar field generated by $B$ due to the motion
of particle $A$, see (\ref{first}). This contribution is not divergent and involves the second derivative of the field $\phi^{(0)}_B$ generated by particle $B$.
Expanding the  Lagrangian for particle $A$ to second order in the velocities and to leading order in $G_N$ and $1/M^4$ we obtain
\begin{eqnarray}
{\cal L}_A&=&\frac{1}{2} m_A(1+\frac{\beta}{m_{\rm Pl}}  \bar \phi^{}_B ( x_A)) \vec v_A^2 -m_A -m_A \Phi_N^B (x_A) - \frac{\beta m_A}{m_{\rm Pl}}  \bar \phi^{}_B ( x_A) +m_A A_i^B v^i_A- \frac{3}{2}  m_A \Phi_N^B (x_A)\vec v_A^2 \nonumber \\
&&+\frac{m_A}{M^4}D_A \nonumber \\
\end{eqnarray}
The terms involving the conformal coupling  renormalise the kinetic energy and the potential energy of the particle.

Let us focus first on the terms coming from the kinetic energy and the Newtonian potential only. We get for the two particles
\begin{eqnarray}
&&{\cal L}_{\rm matter}\supset \frac{1}{2} m_A \vec v_A^2 +\frac{1}{2} m_B \vec v_B^2  -m_A-m_B  -m_A \Phi_N^B (x_A)-m_B \Phi_N^A (x_B)
\nonumber \\ &&
- \frac{3}{2}  m_A \Phi_N^B (x_A)\vec v_A^2-\frac{3}{2}  m_B \Phi_N^A (x_B)\vec v_B^2 -\frac{8 G_N m_A m_B \vec v_A.\vec v_B}{\vert \vec x_A -\vec x_B\vert}. \nonumber \\
\end{eqnarray}
We can add the counter terms $S_{EH}+ S_{GF}$ to obtain the Lagrangian
\be
{\cal L}_{\rm grav}= \frac{1}{2} m_A \vec v_A^2 +\frac{1}{2} m_B \vec v_B^2  -m_A-m_B  +\frac{G_Nm_Am_B}{\vert \vec x_B-\vec x_A\vert }
+ \frac{G_N m_A m_B}{2\vert\vec  x_A -\vec x_B\vert}(3 \vec v_A^2 +3 \vec v_B^2 -8 \vec v_A.\vec v_B + \vec v_{A\perp}.\vec v_{B\perp}).
\ee
We can now add the contributions of the scalar field to this Lagrangian in order to evaluate the effects of both the conformal and disformal interactions.
\subsubsection{The scalar action}
\label{sec:scal}
Here we
 collect all  the scalar field expressions allowing one to complete the action for two moving particles.
The scalar field contributes to the scalar Lagrangian $-\frac{1}{2}(\partial \phi)^2$. After integration by parts and upon using the equation of motion we find
\be
{\cal L}_{\rm scalar}=-\frac{1}{2}\int d^3 x \left( \frac{\beta \phi (x)}{m_{\rm Pl}} (T^A(x) +T^B(x)) +\frac{1}{M^4} \partial_\mu\phi(x) \partial_\nu\phi(x)  (T^{A\mu\nu} +T^{B\mu\nu})(x) \right)
\ee
After expanding
\be
T^{A,B}(x) =- m_{A,B} (1-\frac{\vec v_{A,B}^2}{2}) \delta^{(3)} (\vec x -\vec x_{A,B})
\ee
and using (\ref{trick}) to replace $\frac{\vec v_{A,B\perp}^2}{\vert \vec x_A -\vec x_B\vert }\equiv \frac{\vec v_{A\perp}.\vec v_{B\perp}}{\vert \vec x_A -\vec x_B\vert }$ the first term becomes
\begin{eqnarray}
&&-\frac{1}{2}\int d^3 x  \frac{\beta \phi (x)}{m_{\rm Pl}} (T^A(x) +T^B(x))=\nonumber \\
&&
-\frac{2\beta^2 G_N m_A m_B}{\vert x_B-x_A\vert} (1- \frac{\vec v_A^2}{2} -\frac{\vec v_B^2}{2}+ \frac{\vec v_{A\perp}.\vec v_{B\perp}}{2}) + \frac{\beta}{2m_{\rm Pl}} m_A \delta \phi^{(0)}_{BA}( x_A)+ \frac{\beta}{2m_{\rm Pl}} m_B \delta \phi^{(0)}_{AB}( x_B)\nonumber \\
\end{eqnarray}
where the last term involves the fields from which the self energy divergences have been removed
\be
\delta \phi^{(0)}\equiv \delta\phi^{(0)}_{AB,BA} +{\cal O}(\frac{1}{\vert \vec x-\vec x_{B,A}\vert}).
\ee
Expanding the last term of the action in terms of the regularised field with no self energy divergences $\phi^{({0})}_{A,B}$ in (\ref{corB}) we have finally
 as
\begin{eqnarray}
&&{\cal L}_{\rm scalar}= -\frac{2\beta^2 G_N m_A m_B}{\vert x_B-x_A\vert} (1- \frac{\vec v_A^2}{2} -\frac{\vec v_B^2}{2}+ \frac{\vec v_{A\perp}.\vec v_{B\perp}}{2}) + \frac{\beta}{2m_{\rm Pl}} m_A \delta \phi^{(0)}_{BA}( x_A)+ \frac{\beta}{2m_{\rm Pl}} m_B \delta \phi^{(0)}_{AB}( x_B)\nonumber \\ && - \frac{m_A}{2M^4}D_A -\frac{m_B}{2M^4}D_B \nonumber \\
\label{countS}
\end{eqnarray}
which acts as a counter-term preventing any double counting too.
The end result for the scalar Lagrangian when adding the scalar contributions in the  matter action  and the scalar one is
\begin{eqnarray}
&&{\cal L}_{S}= \frac{2\beta^2 G_N m_A m_B}{\vert x_B-x_A\vert}(1+\frac{\vec v_{A\perp}.\vec v_{B.\perp}}{2}) - \frac{\beta}{2m_{\rm Pl}} m_A \delta \phi^{(0)}_{BA}(\vec x_A)- \frac{\beta}{2m_{\rm Pl}} m_B \delta \phi^{(0)}_{AB}(\vec x_B)\nonumber \\
&&+\frac{\beta}{2m_{\rm Pl}}m_A \phi^{(0)}_B (\vec x_A)) v_A^2 +\frac{\beta}{2m_{\rm Pl}}m_B\phi^{(0)}_A (\vec x_B)) v_B^2 \nonumber \\
 && + \frac{m_A}{2M^4}D_A + \frac{m_B}{2M^4} D_B .\nonumber \\
\end{eqnarray}
This can be explicitly evaluated and gives
\begin{eqnarray}
&&{\cal L}_{S}=-\frac{\beta^2 G_N m_A m_B}{\vert x_B-x_A\vert}(\vec v_A^2 +\vec v_B^2)  +\frac{2\beta^2 G_N m_A m_B}{\vert x_B-x_A\vert}(1+\frac{\vec v_{A\perp}.\vec v_{B.\perp}}{2})\nonumber \\
&&+\frac{\beta^2 G_N}{4\pi}m_Am_B (m_A+m_B)  \frac{((\vec v_A-\vec v_B)_\perp)^2- ((\vec v_A-\vec v_B).\vec n_{AB})^2}{M^4\vert x_A -x_B\vert^4}\nonumber \\
\end{eqnarray}
This contributes to the effective dynamics of the two body system.

\subsubsection{The complete action}
The complete Lagrangian for the two body system combines both the gravitational Lagrangian ${\cal L}_{\rm grav}$  and the contribution from the scalars ${\cal L}_S$
\begin{eqnarray}
&& {\cal L}_{AB}\equiv {\cal L}_{\rm grav}+ {\cal L}_S=
\frac{1}{2} m_A \vec v_A^2 +\frac{1}{2} m_B \vec v_B^2  -m_A-m_B  +\frac{G_N(1+2\beta^2) m_Am_B}{\vert \vec x_B-\vec x_A\vert}\nonumber \\
&& + \frac{G_N m_A m_B}{2\vert \vec x_A -\vec x_B\vert}((3-2\beta^2) \vec v_A^2 +(3-2\beta^2) \vec v_B^2 -8 \vec v_A.\vec v_B + (1+2\beta^2) \vec v_{A\perp}.\vec v_{B\perp})\nonumber\\
 && +\frac{\beta^2 G_N}{4\pi}m_Am_B (m_A+m_B)  \frac{((\vec v_A-\vec v_B)_\perp)^2- ((\vec v_A-\vec v_B).\vec n_{AB})^2}{M^4\vert \vec x_A -\vec x_B\vert^4}.\nonumber\\
\label{uii}
\end{eqnarray}
This Lagrangian is all that is required to obtain the effective metric in the centre of mass frame at leading order in $G_N$ and $1/M^4$.

\subsection{The role of counter terms}
\label{sec:coun}

The calculations of the previous  section have been carried out for two particles interacting both gravitationally and via a scalar field. The Lagrangian for a two body system  has been obtained in several steps for which the role of counter term played by the gravitational and scalar actions is crucial. For the
gravitational and scalar interaction mediated by the conformal coupling, the counter terms serve  only as  book keeping devices which ensure that double counting does not occur.
Let us illustrate this with the case of  two bodies $A$ and $B$ with non-relativistic velocities $\vec v_{A,B}$. At leading order the scalar field is the sum of two contributions, each sourced by the
mass of the particles,  so we find
\be
\phi(x)= \phi_A^{({0})} (x) + \phi_B^{({0})} (x)
\ee
where
\be
\phi_A^{({0})} (x)\sim -\frac{\beta m_{A,B}}{4\pi m_{\rm Pl} \vert \vec x -\vec x_{A,B}\vert}.
\ee
where we have dropped the velocity dependent parts as we are here only interested in the static interaction potential between the bodies.
The contributions to the matter Lagrangian  from this scalar field read simply
\be
{\cal L}_{AB} \supset - m_A \frac{\beta \phi_B^{({0})} (x_A)}{m_{\rm Pl}}-  m_B \frac{\beta \phi_A^{({0})} (x_B)}{m_{\rm Pl}}= \frac{4 \beta^2 G_N m_A m_B}{ \vert \vec x_A -\vec x_B\vert}
\ee
where we have removed the divergent self-energy parts. Notice that this is twice the interaction potential between particles $A$ and $B$. This double counting which  occurs as we have added the matter actions
for both particles is in fact absent as the scalar field Lagrangian $-\frac{1}{2}(\partial \phi)^2$ gives a counter term
\be
\Delta {\cal L}_{AB} \supset  m_A \frac{\beta \phi_B^{({0})} (x_A)}{2m_{\rm Pl}}+ m_B \frac{\beta \phi_A^{({0})} (x_B)}{2m_{\rm Pl}}
\ee
such that the overall Lagrangian only contains one copy of the interaction potential. The same compensation occurs for the Newtonian potential between the two particles.
Notice that one could have used a "symmetrisation" principle and obtain the same result by taking the interaction potential obtained from the action of particle $A$ and then,  realising
that it is symmetric in $A\to B$,  inferred that this must be the actual interaction potential between the two particles. We could have also symmetrise the result from the action of particle $A$ by adding a contribution for which $A\to B$ and dividing the overall result by two. The proper and unambiguous way of obtaining the interaction potential is the one we have outlined, i.e. calculating both the matter, gravity and scalar actions.

For the disformal coupling the matter action for particles $A$ and $B$ involves four contributions which depend on the disformal coupling.
The first two come from the fact that the moving particles source the scalar field in a velocity-dependent way leading to two terms
\be
{\cal L}_{AB}\supset -\frac{\beta}{m_{\rm Pl}} m_A \delta \phi^{(0)}_{BA}(x_A)- \frac{\beta}{m_{\rm Pl}} m_B \delta \phi^{(0)}_{AB}(x_B)
\ee
where the self-energy parts have been removed. Notice that the two contributions involve different combinations of the masses, i.e. respectively $m_A^2 m_B$ and $m_B^2 m_A$. Moreover
they are Galilean invariant and involve only the difference $\vec v_A -\vec v_B$.
The disformal part of the Jordan metric leads to two other terms
\be
{\cal L}_{AB}\supset -\frac{m_A}{M^4} (\partial_\mu \phi_B^{(0)} (x_A) v^\mu_A)^2 -\frac{m_B}{M^4} (\partial_\mu \phi_A^{(0)} (x_B) v^\mu_B)^2
\ee
where $v^\mu_{A,B}=(1,\vec v_{A,B})$ at this order and we  have $\partial_0 \phi_{A,B}^{(0)}=- \vec \partial \phi_{A,B}^{(0)}. \vec v_{A,B}.$ The scalar field
action plays the same role as in the gravitational and conformal cases and simply removes half of the previous terms
\be
\Delta {\cal L}_{AB}= \frac{\beta}{2m_{\rm Pl}} m_A \delta \phi^{(0)}_{BA}(x_A)+\frac{\beta}{2m_{\rm Pl}} m_B \delta \phi^{(0)}_{AB}(x_B)+\frac{m_A}{2M^4} (\partial_\mu \phi_B^{(0)} (x_A) v^\mu_A)^2 +\frac{m_B}{2M^4} (\partial_\mu \phi_A^{(0)} (x_B) v^\mu_B)^2.
\ee
Overall the disformal contributions to the Lagrangian can be combined into two pairs
\be
- \frac{\beta}{2m_{\rm Pl}} m_A \delta \phi^{(0)}_{BA}(x_A)+\frac{m_B}{2M^4} (\partial_\mu \phi_A^{(0)} (x_B) v^\mu_B)^2  =\frac{\beta^2 G_N}{4\pi}m_A^2m_B   \frac{((v_A-v_B)_\perp)^2}{M^4\vert x_A -x_B\vert^4}
\label{pair}
\ee
where
\be
\frac{m_B}{2M^4} (\partial_\mu \phi_A^{(0)} (x_B) v^\mu_B)^2=\frac{m_B}{2M^4} D_B
\ee
and symmetrically for $A\to B$. It is important to notice that the two pairs have different origins. The term $-\frac{\beta}{2m_{\rm Pl}} m_A \delta \phi^{(0)}_{BA}(x_A)$ comes from the matter action of particle $A$
and the corresponding counter term. The term $\frac{m_B}{2M^4} (\partial_\mu \phi_A^{(0)} (x_B) v^\mu_B)^2$ comes from the matter action for $B$ and its counter term. It turns out that the contribution from
the matter action for $A$ combines with the  term from the matter action for $B$, and vice versa.

This implies that one cannot isolate the action for either $A$ or  $B$ in order to investigate the dynamics of the two body system. This  is particular to the disformal interaction compared to the gravitational and conformal ones. In hindsight one could have taken the action for particle $A$ minus the associated counter term and obtained the correct action by symmetrising the result in $A\to B$, i.e. by taking half the sum of the Lagrangian. On the contrary if one only selected the leading term in $m_A m_B^2$ in the matter action for $A$ one would wrongly omit the term in $m_A m_B^2$ coming from the action for $B$ which combine pairwise as in (\ref{pair}). Overall, a much more straightforward way of obtaining the complete action for the two body system with the disformal interaction is to calculate the matter, gravitational and scalar actions as we have done.

\subsection{Centre of mass dynamics}

The previous Lagrangian (\ref{uii}) involving the two bodies $A$ and $B$ can be projected onto a single particle Lagrangian by going to the centre of mass frame.
We will do this by first introducing the Newtonian centre of mass frame coordinates
\be
\vec X= \frac{m_A \vec x_A + m_B \vec x_B}{m_A +m_B}, \ \ \vec x = \vec x_A -\vec x_B
\ee
from which we get the velocities
\be
\vec v_{A,B}= \vec V +\frac{\mu}{m_{A,B}} \vec v.
\ee
The total Lagrangian becomes the sum of a free Lagrangian ($x=\vert \vec x\vert$)
\be
{\cal L}_0= \frac{1}{2} \mu \vec v^2 + \frac{1}{2} {\cal M} \vec V^2 +\frac{G_N m_A m_B}{x}
\ee
where the reduced mass is
\be
\mu= \frac{m_A m_B}{m_A+m_B}
\ee
and the total mass ${\cal M}= m_A +m_B$.
The interaction Lagrangian is
\begin{eqnarray}
&&{\cal L}_{\rm int}= \frac{G_N m_A m_B}{2 x} (-(1+2\beta^2)\vec V^2 + ((3-2\beta^2)\mu^{-2} + (1+2\beta^2) m_A^{-1} m_B^{-1}) \mu^2 \vec v^2 - (1+2\beta^2) (m_A^{-1}\nonumber\\ && -m_B^{-1}) \mu \vec v.\vec V
-(1+2\beta^2) ( (\vec V.\vec  n)^2 -\frac{\mu^2}{m_A m_B} (m_A^{-1}-m_B^{-1}) ( \vec v. \vec n)^2 +\mu ( \vec v.\vec n) (\vec  V.\vec n)))\nonumber \\
\end{eqnarray}
together with the disformal term
\be
{\cal L}_{\rm dis}= \frac{\beta^2 G_N}{4\pi M^4 x^4 } \mu {\cal M}^2\left ( \vec v_\perp ^2- (\vec v.\vec n)^2\right )
\ee
We have identified $\vec n= \vec n_{AB}$ here.
The absence of $\vec X$ dependence implies that
\be
\vec P= \frac{\partial \cal L}{\partial \vec V}
\ee
is conserved. We set the centre of mass momentum to zero $\vec P=0$ and integrate out $\vec V$ at leading order in $G_N$
\be
{\cal M} \vec V= \frac{G_N (m_A -m_B) \mu}{2 x}  (1+2\beta^2)(\vec v + ( \vec v.\vec n)\vec n) .
\ee
The effective Lagrangian for the velocity $\vec v$ is then given by
\begin{eqnarray}
&&{\cal L}_{\rm eff}= \frac{1}{2} \mu \vec v^2 +\frac{G_N \mu {\cal M}(1+2\beta^2)}{x}+ \frac{G_N \mu {\cal M}}{2x} ((3-2\beta^2) \vec v^2 + (1+2\beta^2) \nu \vec v^2 + (1+2\beta^2) \nu ( \vec v.\vec n)^2)\nonumber \\
&&+\frac{\beta^2 G_N (\vec v_\perp^2-(\vec v.\vec n)^2)}{4\pi} \frac{\mu {\cal M}^2 }{x^4 M^4}\nonumber \\
\end{eqnarray}
where we have introduced the parameter
\be
\nu=\frac{m_A m_B}{(m_A+m_B)^2}.
\ee
Using the identity, at leading order in $G_N$,
\be
\frac{d}{dt}( \vec v. \vec n)\equiv \frac{\vec v^2 -( \vec v.\vec  n)^2}{ x}
\ee
the Lagrangian becomes equivalent to
\be
{\cal L}_{\rm eff}= \frac{1}{2} \mu \vec v^2 +\frac{G_N \mu {\cal M}(1+2\beta^2)}{x}+ \frac{G_N \mu {\cal M}}{2x} ((3-2\beta^2) \vec v^2 + 2(1+2\beta^2) \nu \vec v^2 )
+\frac{\beta^2 G_N (v_\perp^2-(\vec v.\vec n)^2)}{4\pi} \frac{\mu {\cal M}^2 }{x^4 M^4}.
\ee
Let us introduce the effective metric
\begin{eqnarray}
&& g_{00}^{\rm eff}= -(1-\frac{2G_N {\cal M} (1+2 \beta^2)}{x})\nonumber \\
&& g_{ij}^{\rm eff}= (1+ \frac{2G_N {\cal M} (1-2 \beta^2)}{x})\delta_{ij} + \frac{\beta^2 G_N}{2\pi} \frac{ {\cal M}^2 }{x^4 M^4}(\delta_{ij}-2n_i n_j)
\nonumber \\
\label{effe}
\end{eqnarray}
and the reduced Lagrangian
\be
{\cal L}_{\rm red}= - g_{\mu\nu}^{\rm eff} v^\mu v^\nu
\ee
where $v^{\mu}= (1,v^i)$.  Then
we have to leading order that the centre of mass Lagrangian can be reconstructed using
\be
{\cal L}_{\rm eff}= -\mu \sqrt{ {\cal L}_{\rm red} + 2 \nu ({\cal L}_{\rm red}-1)^2}.
\ee
The first term is the Lagrangian of a particle of mass $\mu$ subject to the effective metric created by the mass ${\cal M}$. The quadratic correction in the square root in
 $\nu$  is due to the fact that the masses are not light masses. The extrema of ${\cal L_{\rm eff}}$ can be obtained
by extremising ${\cal L}_{\rm red}$ which depends on the effective metric $g^{\rm eff}_{\mu\nu}$.

The effective metric (\ref{effe}) is known to provide an exact result in the  post-Minkowskian limit \cite{Damour:2016gwp}, i.e. at leading order in $G_N$, and to all order in the velocity in the conformal case. Here we have retrieved this result
using the low $\vec v^2$ expansion and we have extended it to include the first correction in $1/M^4$ due to the disformal coupling. Moreover the derivation of the effective metric is usually carried out in the Hamiltonian formalism
whereas we have worked at the Lagrangian level and at the lowest order in $\vec v^2$ as it sufficient to reconstruct the effective metric.

Notice that the disformal correction has been assumed throughout to be the leading correction to the Newtonian case implying that we can consider this effective metric in situations where
\be
\frac{{\cal M}}{x^3} \lesssim  \frac{M^4}{\beta^2} \lesssim  \frac{m_{\rm Pl}^2}{x^2}.
\ee
If the first inequality were  nearly saturated then Newtonian orbits would be largely affected whereas if the second inequality were violated we would have to take into account the higher order corrections to the metric
in GR.
Taking as an example the orbit of Mercury at an average distance of $6\times 10^7$ km from the sun this leads to
\be
2\times 10^{-3} \ {\rm MeV} \lesssim \frac{M}{\sqrt \beta} \lesssim 7.5 \times 10^{-2}\ {\rm MeV}
\ee
for the disformal interaction to play a relevant role.
As the Cassini bound \cite{Bertotti:2003rm} leads to  $\beta^2 \lesssim  10^{-5}$, this implies that gravitational effects of the disformal coupling could be  relevant for planetary orbits when
\be
M\lesssim 4\times 10^{-3} {\rm MeV}.
\ee
Of course the disformal interaction becomes  relevant for larger values of $M$  in situations where the Newtonian potential is larger, such as the orbits of two neutron stars in their inspiralling phase where objects of masses similar to the sun's orbit at a few hundreds of kilometres from each other. We will comment on this case below.

\section{The dynamics of a light particle}
\label{sec:dyn1}

{ In this section we focus on the dynamics of a light particle and in particular the classical tests of General Relativity such as the Shapiro time delay and the perihelion advance.
In the following we neglect the influence of the cosmological background which would result in a time variation of masses due to the conformal factor $A(\phi_{\rm cosmo})$. We focus on the effects
due to the disformal coupling. The effects of the time drift of masses, or equivalently Newton's constant, in Brans-Dicke theories are well documented as can be found in \cite{Uzan:2002vq}. Typically
the relative variation of masses should be less than one percent of the Hubble rate. }

\subsection{Violation of the  equivalence principle}

In this section we focus on the light particle case which can be obtained from the two body analysis by setting $\nu \to 0$. In this case, the light particle of mass $m_B \ll m_A$ evolves with the dynamical Lagrangian
\be
{\cal L}_{\rm eff}= -m_B \sqrt{ {\cal L}_{\rm red}}.
\label{test}
\ee
where the effective metric is given by
\begin{eqnarray}
&& g_{00}^{\rm eff}= -(1-\frac{2G_N {m_A} (1+2 \beta^2)}{x})\nonumber \\
&& g_{ij}^{\rm eff}= (1+ \frac{2G_N {m_A} (1-2 \beta^2)}{x})\delta_{ij} + \frac{\beta^2 G_N}{2\pi} \frac{ {m_A}^2 }{x^4 M^4}(\delta_{ij}-2n_i n_j)
\nonumber \\
\label{effe1}
\end{eqnarray}
The effective action is the one of a particle evolving in the background metric given by (\ref{effe1}). Notice that the disformal part involves both the perpendicular and parallel velocities. This
is different from the trajectories of photons which follow the null trajectories of the Jordan metric
\be
ds^2_J= -g_{\mu\nu} dx^\mu dx^\nu\equiv 0
\ee
where the Jordan metric is given by
\begin{eqnarray}
&& g_{00}= -(1-\frac{2G_N {m_A} (1+2 \beta^2)}{x})\nonumber \\
&& g_{ij}= (1+ \frac{2G_N {m_A} (1-2 \beta^2)}{x})\delta_{ij} + \frac{\beta^2 G_N}{\pi} \frac{ {m_A}^2 }{x^4 M^4}n_i n_j.
\nonumber \\
\label{effe2}
\end{eqnarray}
which involves the parallel velocity only.
As a result the equivalence principle is violated between photons and matter. This follows from the fact that for non-relativistic matter particles, the mass of a light particle cannot be
neglected completely and generates a field contribution $\delta \phi^{(0)}_{AB}$ whose presence in the matter action of the massive particle is of the same order as the disformal terms in the
matter action of the light particle, see the discussion in section \ref{sec:coun}. We will analyse what differences this induces in both the Shapiro effect, i.e. the time delay of photons, and the perihelion advance, i.e. the motion of a light particle.

\subsection{Shapiro time delay}

The study of the time delay of radio waves compared to its counterpart in GR is crucial as it gives direct access to modifications of GR in the environment of a massive object $A$, typically the sun.
It is convenient to introduce the metric potential $\Phi (r)= \Phi_N (r) + \beta \frac{\phi^{(0)}(r)}{m_{\rm Pl}}$ such that
\be
\Phi (r)= -\frac{G_{\rm eff} m_A}{r}.
\ee
where the effective Newton constant is here
\be
G_{\rm eff}= (1+2\beta^2) G_N.
\ee
We are interested in the time delay of signals sent between two points such that to leading order the photon trajectory is a straight line with an impact parameter $b$, i.e. in terms of polar coordinates
\be
r=\frac{b}{\cos\theta}.
\ee
Along this trajectory the time delay compared to GR is due to the corrections to the metric felt by the photons. This reads
\be
ds^2_J= -(1+2\Phi) dt^2 + (1- 2(1+ 2\gamma) \Phi + \sin^2\theta \cos^4 \theta\frac{G_N\beta^2 m_A^2}{\pi M^4b^4}) d x^2
\ee
where $dx^2= dr^2 +r^2 d\theta^2= \frac{b^2}{\cos^4 \theta} d\theta^2$ and $dr^2=\sin^2 \theta dx^2$.  The last contribution to $dx^2$ comes from the disformal interaction.
We have also introduced the parameter
\be
\gamma=- \frac{2\beta^2}{1+2\beta^2}.
\ee
Let us apply this result to the trajectory of photons between two points $C$ and $D$ which can be taken to be the location of the earth and of Cassini satellite \cite{Bertotti:2003rm}. We have therefore
\be
\frac{dt}{dx}= 1- 2(1+\gamma) \Phi + \sin^2 \theta \cos^4 \theta\frac{G_N\beta^2 m_A^2}{2\pi M^4b^4}
\ee
along the photon trajectory.
The time delay due the modification of gravity is
\be
\frac{d\delta t}{dx}= -2\gamma\Phi + \sin^2\theta \cos^4\theta \frac{G_N\beta^2 m_A^2}{2\pi M^4b^4}
\ee
 such that $\Phi=-\frac{G_{\rm eff}m_A}{b} \cos \theta$. Using $dx= \frac{b}{\cos^2 \theta} d\theta$, this implies that
\be
\frac{d\delta t}{d\theta}=2 \gamma G_{\rm eff} \frac{m_A}{\cos \theta}+ \sin^2\theta \cos^2\theta \frac{G_N\beta^2 m_A^2}{2\pi M^4b^3}.
\ee
Let us assume that the two massive bodies where the signal is received and emitted follow circular trajectories, for simplicity,  around the massive body. This implies that for instance $\cos \theta_C= b/r_C$ and therefore a variation of the position of the emitter or the receiver by $d\theta_{C,D}$ corresponds to a change of impact parameter $ db = -r_{C,D} \sin \theta_{C,D} d\theta_{C,D}$. As a result the variation of the time delay due to  a variation of the impact parameter  is
\be
\frac{d\delta t}{d b}=- 2\gamma G_{\rm eff} \frac{m_A}{b}(\frac{1}{\sin\theta_C}+\frac{1}{\sin\theta_D})- {\sin\theta_C \cos^3\theta_C} \frac{G_N\beta^2 m_A^2}{2\pi M^4b^4}-{ \sin\theta_D\cos^3\theta_D}
\frac{G_N\beta^2 m_A^2}{2\pi M^4b^4}.
\ee
The  probes such as Cassini  which are used to investigate  the time delay for radiowaves  are  going behind the sun for $\theta_{C,D} \sim \frac{\pi}{2}$. As a result the contribution to the time delay from the disformal interaction is negligible. The time delay as measured between two positions of the emitter, after a round trip, with impact parameters $b_1$ and $b_2$ is therefore given by
\be
\delta t_1 -\delta t_2= -8 \gamma G_{\rm eff} m_A \ln \frac{b_1}{b_2}.
\ee
and does not depend on the disformal interaction.

\subsection{The perihelion advance}

One of the classical tests of General Relativity is the advance of perihelion of mercury. Here we will calculate the advance of perihelion for a light particle around a heavy body when
the conformal and disformal interactions are present.
The study of the dynamics of such a light object is  easier to carry out going back to the original Lagrangian (\ref{effe1}).
In particular, we consider that time and space are parameterised in proper time $\tau_B$. The trajectories of massive objects are such that
\be
g^{\rm{eff}}_{\mu\nu}u^\mu_B u^\nu_B=-1
\label{cv}
\ee
where
\be
u^{\mu}_B= \frac{dx^\mu}{d\tau_B}.
\ee
In this section we put $\tau=\tau_B$ and define $\dot \  =d/d\tau$.
In polar coordinates, in the orbital plane, and using
\be
d\Omega^2= d\theta^2
\ee
we have
\begin{eqnarray}
\label{jod}
g_{00}^{\rm eff}&=& -1+ 2\frac{G_N m_A(1+2\beta^2)}{r}\nonumber \\
 g_{rr}^{\rm eff}&=& 1+2\frac{G_N m_A(1-2\beta^2)}{r}-\frac{G_N\beta^2 m_A^2}{2\pi M^4r^4} \nonumber \\
  g_{\theta\theta}^{\rm eff}&=&r^2\left (1+2\frac{G_N m_A(1-2\beta^2)}{r}+\frac{G_N\beta^2 m_A^2}{2\pi M^4r^4}\right ) .\nonumber \\
\end{eqnarray}
Notice that the disformal contribution appears both in the radial and tangential parts of the metric.
As the Lagrangian is independent of $\theta$, the angular momentum $J$ is conserved implying that
\be
r^2(1+2\frac{G_N m_A(1-2\beta^2)}{r}+\frac{G_N\beta^2 m_A^2}{2\pi M^4r^4}) \dot \theta= \frac{J}{m_B}.
\ee
Similarly the absence of any explicit time dependence in $x_0$ implies that
\be
(1- 2\frac{G_N m_A(1+2\beta^2)}{r})\dot x^0= k
\ee
where $k$ is a constant.
The constraint (\ref{cv}) implies that
\begin{eqnarray}
&&\frac{k^2}{(1- 2\frac{G_N m_A(1+2\beta^2)}{r})}  -(1+2\frac{G_N m_A(1-2\beta^2)}{r}-\frac{G_N\beta^2 m_A^2}{2\pi M^4r^4})\dot r^2  \nonumber \\ && -\frac{J^2}{m_B^2}\frac{1}{r^2(1+2\frac{G_N m_A(1-2\beta^2)}{r} +\frac{G_N\beta^2 m_A^2}{2\pi M^4r^4})}=1\nonumber \\
\end{eqnarray}
The dynamics are most conveniently analysed by changing coordinates and introducing the spherical distance
\be
\tilde r^2 =(1+ \frac{2G_N m_A }{r}(1-2\beta^2))r^2
\ee
which corresponds to writing the angular part of the metric as
$g_{\theta\theta}=\tilde r^2$ in the absence of disformal interaction.
We obtain that to leading order in $G_N$
\be
\tilde r= r+{G_N m_A }(1-2\beta^2).
\ee
and $\dot{\tilde r}= \dot r$.
This implies that angular momentum conservation can be reformulated as
\be
\tilde r^2(1+\frac{G_N\beta^2 m_A^2}{2\pi M^4\tilde r^4})  \dot \theta= \frac{J}{m_B}.
\ee
At leading order in $G_N$ and reverting to $\tilde r \to r$ for convenience we have now
\be
\frac{k^2}{(1- 2\frac{G_N m_A(1+2\beta^2)}{r})} -(1+2\frac{G_N m_A(1-2\beta^2)}{r}-\frac{G_N\beta^2 m_A^2}{2\pi M^4r^4})\dot r^2 - \frac{J^2}{m_B^2r^2(1+\frac{G_N\beta^2 m_A^2}{2\pi M^4r^4})}=1.
\ee
Let us now introduce the Binet variable $u=1/r$ such that
\be
\dot r= -\frac{J}{m_B} \frac{du}{d\theta}.
\ee
We then obtain the following differential equation
\begin{eqnarray}
(\frac{du}{d\theta})^2(1- 8 G_N \beta^2 m_A u-\frac{G_N\beta^2 m_A^2}{2\pi M^4}u^4) +u^2 = && \frac{k^2-1}{J^2}m_B^2 + \frac{2G_N (1+2\beta^2) m_Am_B^2}{J^2} u\nonumber \\ && +2G_N m_A (1+2\beta^2) u^3 +\frac{G_N\beta^2 m_A^2}{2\pi M^4}u^6\nonumber \\
\end{eqnarray}
This is the main equation for the dynamics of planar orbits involving both conformal and disformal interactions.

By taking the derivative of the previous relation  we deduce the generalised Binet equation
\be
\frac{d^2u}{d\theta^2}+u= (4 G_N \beta^2 m_A+\frac{G_N\beta^2 m_A^2}{\pi M^4}u^3) (\frac{du}{d\theta})^2 + \frac{G_N(1+2\beta^2) m_Am_B^2}{J^2} + G_Nm_A (3 -2\beta^2) u^2 +\frac{G_N \beta^2 m_A^2}{\pi M^4} u^5.
\label{binet}
\ee
which reduces to the one in General Relativity when $\beta=0$. The new terms due to the conformal and disformal interactions modify the structure of the orbits.

 One can construct solutions
 in perturbation theory around the classical trajectory
\be
u_0= \frac{G_N(1+2\beta^2) m_Am_B^2}{J^2} (1+ e \cos \theta)
\ee
where the semi long-axis is
\be
a= \frac{J^2}{m_B^2 m_A G_N(1+2\beta^2)} \frac{1}{1-e^2}
\ee
corresponding to
\be
u_0= \frac{1}{a(1-e^2)} (1+ e\cos \theta).
\ee
The first correction to the classical trajectory satisfies
\be
\frac{d^2u_1}{d\theta^2}+u_1= 4\pi G_N \beta^2 m_A (\frac{du_0}{d\theta})^2 +  G_Nm_A (3 -2\beta^2) u_0^2 +\frac{G_N \beta^2 m_A^2}{\pi M^4} u_0^5.
\label{bin}
\ee
Notice that the source terms are all proportional to $G_N$ as befitting our expansion scheme.

We will not solve this equation in full generality. As we are only interested in the advance of perihelion, we select the source terms on the right hand side of the perturbed Binet equation
(\ref{bin}) in $\cos\theta$. Higher harmonics are present and will not give rise to contributions to the advance of perihelion. As a result we only need to select the $\cos\theta$ source terms which correspond to
\be
\frac{d^2u_1}{d\theta^2}+u_1\supset \left(2\frac{e}{a^2(1-e^2)^2 }G_Nm_A (3 -2\beta^2)  +\frac{e}{a^5(1-e^2)^5}\frac{5G_N \beta^2 m_A^2}{\pi M^4}\right ) \cos \theta .
\label{bine}
\ee
whose solution is
\be
u_1= \alpha \theta \sin \theta
\ee
with
\be
\alpha= \left(\frac{e}{a^2(1-e^2)^2 }G_Nm_A (3 -2\beta^2)  +5\frac{e}{a^5(1-e^2)^5}\frac{G_N \beta^2 m_A^2}{2\pi M^4}\right ).
\ee
As a result we have at this order
\be
u=u_0+u_1 \equiv \frac{1}{a(1-e^2)} (1+ e\cos\left( (1-\frac{\alpha a}{e}(1-e^2))\theta\right))
\ee
and therefore the perihelion advance is given by
\be
\Delta \theta= \frac{2\pi \alpha a}{e}(1-e^2)
\ee
or more directly
\be
\Delta \theta=2\pi \frac{G_N m_A}{p}\left ( (3 -2\beta^2)  +5\frac{ \beta^2 m_A}{2\pi M^4p^3}\right )
\ee
where we have introduced
\be
p= a(1-e^2).
\ee
The perihelion advance can be written as
\be
\Delta \theta=2\pi \frac{G_{\rm eff} m_A}{p}\left ( 3+ \gamma(4  -\frac{ 5 m_A}{4\pi M^4p^3})\right ).
\ee
The first term is the result in GR corrected by the scalar-tensor coupling \cite{Will:2004nx}, the last term is new and comes from the disformal interaction.
Notice that the GR and conformal cases have been retrieved whilst never going beyond the leading $G_N$ corrections.

As the overall result depends on both $\beta$ and $M$, no precise bound on $M$ can be deduced. A reasonable requirement may be
\be
M^4\gtrsim \frac{m_\odot}{ p^{3}}
\ee
for planets orbiting around the sun. This is of course not mandatory as $\beta$ might be very small. For Mercury this would simply require that
\be
M\gtrsim 10^{-4} \ {\rm MeV}
\ee
which is weaker than the E\"otwash bound $M\ge 0.07 $ MeV \cite{Brax:2014vva}.

\section{Discussion and Conclusions}
\label{sec:con}
We have analysed the dynamics of bodies interacting both gravitationally and via a scalar field which can be exchanged between moving objects. We have seen that the disformal coupling has an effect
only when combined with a conformal interaction. In this case, the disformal coupling leads to a change of the advance of perihelion for a light body and modifies the effective metric
which governs the evolution of two interacting bodies. We have shown that contrary to GR and conformal couplings for which the advance of perihelion is proportional to the mass
of the heavy object around which a light particle orbits, the disformal coupling leads to a quadratic dependence. Although we have not considered the case of black holes, as in particular the no-hair theorem implies that
no scalar field is generated outside the horizon, we can certainly envisage that for astrophysical black holes of several million solar masses and with accretion discs \cite{Davis:2016avf,Davis:2014tea}, a large scalar field would be generated and therefore
we expect that because of the large mass of the black holes, there could be  large effects on the dynamics of stars in the vicinity of the centre of a galaxy like the Milky-Way. It would be worth analysing
this possibility and setting bounds on the disformal coupling from the advance of perihelion of such stars orbiting the galactic centre.
In this paper we have also shown that for a two body system with conformal and disformal interactions, the centre of mass dynamics can be captured by an effective metric at leading
order in $G_N$.  The effects of the
disformal coupling in the case of two inspiralling neutron stars could be relevant for future observations and would give us indications on the existence of both conformal and disformal interactions
between matter and a scalar field.
We expect that the disformal interaction begins to induce large deviations from General Relativity when
\be
M^4 \lesssim  \frac{\beta ^2 m_\odot}{ R^{3}}
\ee
for objects of masses around one solar mass at a distance $R$, see (\ref{binet}). Typically the Cassini experiment implies that $\beta^2 \lesssim 10^{-5}$ \cite{Bertotti:2003rm}, and therefore the typical order
of magnitude of the upper bound for which disformal effects might be expected when two neutron stars are separated by $R\sim 100$ km is
\be
M\lesssim 3\ {\rm MeV}.
\ee
This has to be compared with the constraints on disformal couplings in different environments (see Table 1 in \cite{Brax:2014vva}), i.e. the scale $M$ could be different for physical processes involving various densities or energy scales.
Typically such a range for $M$ is compatible with  the lower bound on $M\ge 0.07 $ MeV from the E\"otwash experiment \cite{Brax:2014vva}. In conclusion, we find that disformal effects could play a role in the merging
of two neutron stars when the disformal scale is $M\sim 1$ MeV. The details of this study are left for the future.

\section*{Acknowledgments}

We would like to thank F\'elix Juli\'e for discussions and suggestions. ACD  acknowledges partial support from STFC under grants ST/L000385/1 and ST/L000636/1.
This work is supported in part by the EU Horizon 2020 research and innovation programme under the Marie-Sklodowska grant No. 690575. This article is based upon work related to the COST Action CA15117 (CANTATA) supported by COST (European Cooperation in Science and Technology).

\bibliographystyle{JHEP}
\bibliography{ref1}

\end{document}